\documentstyle[twocolumn,aps,epsfig]{revtex}  
\begin{document} 
\draft           
\title{Klein Paradox for Optical Scattering from Excited Targets}
\author{O. Panella}
\address{Istituto Nazionale di Fisica Nucleare, Sezione di Perugia, 
Via A. Pascoli, Perugia Italy}
\author{Y.N. Srivastava and A. Widom}  
\address{Physics Department \& INFN, University of Perugia, Perugia Italy}
\address{Physics Department, Northeastern University, Boston MA USA}
\maketitle

\begin{abstract}                
The well known Klein paradox for the relativistic Dirac wave equation 
consists in the computation of possible ``negative probabilities'' induced 
by certain potentials in some regimes of energy. The paradox may be resolved 
employing the notion of electron-positron pair production in which the 
number of electrons present in a process can increase. The Klein paradox 
also exists in Maxwell's equations viewed as the wave equation for photons. 
In a medium containing ``inverted energy populations'' of excited atoms, 
e.g. in a LASER medium, one may again compute possible ``negative 
probabilities''. The resolution of the electromagnetic Klein paradox is 
that when the atoms decay, the final state may contain more photons then 
were contained the initial state. The optical theorem total cross section 
for scattering photons from excited state atoms may then be computed as 
negative within a frequency band  with matter induced amplification. 
\end{abstract} 
\narrowtext

\section{Introduction} 

At a time in which the properties of relativistic electrons (implicit 
in the  Dirac equation\cite{1} for the spinor wave function) were not 
so well understood, Klein predicted a ``paradox'' associated with 
the problem of reflecting an electron off a step potential. 
The reflection coefficient 
\begin{math} R \end{math} and the transmission coefficient 
\begin{math} P \end{math} (induced by the step potential) obey an 
expected sum rule 
\begin{equation} 
R+P=1\ \ {\rm  (exact)}. 
\end{equation} 
However, for some step potential heights and for some incident energies  
\begin{equation} 
R>1\ \ {\rm  (possible)}. 
\end{equation} 
Eqs.(1) and (2) imply the possibility of a negative transmission 
``probability'' coefficient; i.e. \begin{math} P<0 \end{math} is possible 
and therein lies the paradox. 

The Dirac wave equation for electrons and positrons allows for a 
physical picture which clarifies the meaning of the Klein paradox\cite{2} 
and thus removes it\cite{3,4,5,6} as a possible objection to the Dirac 
theory. When an electron hits the step potential, it is possible to 
create an electron-positron pair. For such an event, two electrons may 
be reflected from the step potential when only one electron was incident. 
The added positron may then be transmitted into the step potential. 
Such an event, if sufficiently probable, may yield 
\begin{math} R>1 \end{math} since more electrons may be 
reflected from the step potential than are incident. Furthermore, 
\begin{math} P<0 \end{math} may be interpreted to mean that the current 
of the transmitted positron wave is directed oppositely to that 
which would have been carried by an electron wave moving at the same 
velocity.

Our purpose is to point out that the Klein paradox\cite{7,8,9} is 
intrinsic to relativistic particles\cite{10,11,12,13} and in particular 
applies to the photon. For the electromagnetic case, the wave equations 
for the photon in space-time are merely Maxwell's equations. To 
understand the nature of the Klein paradox for photons we may recall 
the classical results discussed by Rayleigh. Suppose a spherically 
symmetric target with a fluctuating dipole moment which scatters an 
electromagnetic wave. We suppose that the electric dipole moment 
\begin{math} {\bf p}(t)  \end{math} 
response to an applied electric field 
\begin{equation} 
{\bf E}(t)={\Re}e\left\{{\bf E}_0 e^{-i\zeta t}\right\} 
\ \ {\rm with}\ \ {\Im}m (\zeta )>0,
\end{equation}
is given by 
\begin{equation} 
{\bf p}(t)={\Re}e\left\{\alpha (\zeta ){\bf E}_0 e^{-i\zeta t}\right\}. 
\end{equation}
Here \begin{math} \alpha (\zeta ) \end{math} is the target 
polarizability. 

Rayleigh asserted that the elastic scattering amplitude for an 
incident electromagnetic wave onto the polarizable target is given 
in the dipole approximation by 
\begin{equation}
F_{i\to f}=\left({\omega \over c}\right)^2 {\bf e}_f^* \cdot {\bf e}_i ,
\alpha (\omega +i0^+)
\end{equation}
where \begin{math} {\bf e}_i  \end{math} and 
\begin{math} {\bf e}_f \end{math} 
are, respectively, the initial and final polarization vectors. 
Averaging over initial polarization and summing over final 
polarizations yields the elastic differential cross section 
\begin{eqnarray}
\left({d\sigma_{el}\over d\Omega }\right)&=&
{1\over 2}\sum_i \sum_f \left|F_{i\to f}\right|^2 \nonumber \\ 
 &=  &
{1\over 2}
\left(1+\cos ^2 \theta \right)\left({\omega \over c}\right)^4
\left|\alpha (\omega +i0^+)\right|^2,
\end{eqnarray}
which implies for the elastic cross section 
\begin{math} \sigma_{el}=\int d\sigma_{el} \end{math} that   
\begin{equation}
\sigma_{el}(\omega )=
\left({8\pi \over 3}\right)\left({\omega \over c}\right)^4
\left|\alpha (\omega +i0^+)\right|^2 .
\end{equation}

For understanding the Klein paradox for electromagnetic waves it is 
more important to discuss the total cross section which involves 
the optical theorem 
\begin{equation}
\sigma_{tot}=\left({4\pi c\over \omega }\right){\Im}m \big(F_{i\to i}\big) 
\end{equation}
in the form 
\begin{equation}
\sigma_{tot}(\omega )=
\left({4\pi \omega \over c}\right)
{\Im }m\left(\alpha (\omega +i0^+)\right).
\end{equation}
The expected sum rule is that 
\begin{equation}
\sigma_{tot}(\omega )=\sigma_{el}(\omega )+\sigma_{in}(\omega ),
\end{equation}
where \begin{math} \sigma_{in}(\omega ) \end{math} is the inelastic 
cross section. 
In the commonly studied case in which the target ``absorbs'' radiation 
one has a positive total cross section 
\begin{math}\sigma_{tot} (\omega )>0\end{math} 
since  
\begin{equation}
\omega {\Im }m\left(\alpha (\omega +i0^+)\right)>0 
\ \ \ {\rm (absorption\  band)}.
\end{equation}
On the other hand, for a target in some excited energy state, (say) with 
atoms having ``inverted' energy populations, there will exist frequency 
bands in which 
\begin{equation}
\omega {\Im }m\left(\alpha (\omega +i0^+)\right)<0 
\ \ \ {\rm (amplifier\ band)}.
\end{equation} 
The optical theorem in the form of Eqs.(9) and (12) yield the following.
\par \noindent
{\bf The Optical Klein Paradox:} If \begin{math}{\cal B}\end{math} is 
the set of frequencies within which the target is an electromagnetic 
amplifier, then  
\begin{equation}
\sigma_{tot}(\omega \in {\cal B})<0.
\end{equation}

A negative total cross section in an amplifying frequency band is only 
at first glance an impossibility. The purpose of this work is to 
discuss physical meaning of the optical Klein Paradox for amplifiers 
such as inverted population targets,  
e.g. ``pumped'' LASER materials\cite{14}. For the case of the 
Dirac equation, pair production allowed (without changing total charge) 
for more electrons in an outgoing state than were present in the 
incoming state. This simple fact makes transparent the notion of 
a negative forward transmission coefficient \begin{math} P<0 \end{math}. 
Similarly, for an excited state target one incident photon can give rise 
to two photons in the outgoing state when the excited state target is 
induced to decay (say) into a target ground state. In an electromagnetic 
scattering experiment, there can be more radiation behind the target than 
that which would exist if the excited radiating target were removed. 

In Sec.II, we consider an electromagnetic wave traveling through a 
medium with a dielectric response function 
\begin{math} \varepsilon(\zeta ) \end{math}.
For a dilute density per unit volume \begin{math} n \end{math} of 
polarizable targets in the medium, the dielectric response 
\begin{equation}
\varepsilon (\zeta )=1+4\pi n\alpha (\zeta )+\ ...\ \  
{\rm if}\ \ |4\pi n\alpha (\zeta )|<<1,
\end{equation} 
so that 
\begin{equation}
{\Im}m \big(\varepsilon (\omega +i0^+)\big) 
= 4\pi n{\Im}m \big(\alpha (\omega +i0^+)\big).
\end{equation}
The properties of 
\begin{math}  \varepsilon (\omega +i0^+) \end{math} within an amplifying 
frequency band \begin{math} {\cal B} \end{math} will be explored. 
In Sec.III, the concept of a negative noise temperature 
\begin{math} T_n \end{math} will be defined. The definition of an 
amplifying target will be related to the notion of negative noise 
temperature \begin{math} T_n \end{math}\cite{15,16} . It will 
be shown that an amplifying frequency band 
\begin{math} {\cal B} \end{math} may equally 
well be defined by the notion of a negative noise temperature 
\begin{math} T_n(\omega \in {\cal B})<0  \end{math}. The electromagnetic 
Klein paradox occurs in all material systems exhibiting a 
negative radiation noise temperature. The general optical theorem will 
be proved in Sec.IV without regard to the sign of the noise temperature. 
Physical examples will be discussed in the concluding Sec.V. 

\section{A Traveling Plane Wave in Matter}

Consider an electromagnetic plane wave 
\begin{equation}
{\bf E}={\Re}e\left({\bf E}_0 e^{i(kz-\omega t)}\right)
\end{equation}
traveling through a medium with a dielectric response function 
\begin{math} \varepsilon(\zeta) \end{math} so that
\begin{equation}   
k={\omega \over c}\sqrt{\varepsilon(\omega +i0^+)}.
\end{equation}
The intensity of the light beam described by the plane wave 
is proportional to 
\begin{equation}
\overline{|{\bf E}|^2}={1\over 2}|{\bf E}_0 |^2 \exp\left(-hz \right)
\end{equation}
defines the extinction coefficient 
\begin{equation}
h=2{\Im}m(k).
\end{equation}
From Eq.(18) it is evident that 
\begin{equation}
\matrix{
{\rm an\ absorbing\ medium}\ \Longrightarrow \ (h>0) \cr 
{\rm and} \cr
{\rm an\ amplifying\ medium}\ \Longrightarrow \ (h<0).  
}
\end{equation}
For a medium consisting of a dilute gas of polarizable particles 
for which \begin{math} n|\alpha (\omega +i0^+)|<<1 \end{math},  
it follows from Eqs.(14), (17) and (19) that 
\begin{equation}
h=n\sigma_{tot}=4\pi (\omega /c){\Im }m\alpha (\omega +i0^+).
\end{equation}
Eqs.(20) and (21) imply the optical Klein theorem: 
\begin{equation}
\matrix{
{\rm an\ absorbing\ medium}\ \Longrightarrow \ (\sigma_{tot}>0) \cr 
{\rm and} \cr
{\rm an\ amplifying\ medium}\ \Longrightarrow \ (\sigma_{tot}<0).  
}
\end{equation}
The paradox of having 
\begin{math} \sigma_{tot}(\omega \in {\cal B})<0 \end{math} 
in an amplifying frequency band \begin{math} {\cal B} \end{math} 
has now been formally proved. 

\section{Noise Temperature and Spectral Functions}

The quantum spectral functions for dipole moment 
\begin{math} {\bf p}(t) \end{math} fluctuations in a target 
may be defined by 
$$
S_\pm (\omega )=
$$
\begin{equation}
{1\over 3}\sum_I\sum_F p_I 
\left|\left<F\left|{\bf p}\right|I \right>\right|^2 
\delta \left(\omega \mp {(E_F-E_I)\over \hbar}\right),
\end{equation}
where \begin{math} p_I \end{math} is the target probability 
of being in an initial energy eigenstate  
\begin{math} H\left|I \right >=E_I\left|I \right> \end{math}.
If the initial target is at temperature \begin{math} T \end{math},  
\begin{equation}
p_I{\rm (Thermal)}=
\exp\left({{\cal F}-E_I\over k_B T}\right)
\ \ {\rm (Equilibrium)},
\end{equation}
then the spectral functions in Eq.(23) obey the detailed balance 
condition
\begin{equation}
S_-(\omega )=S_+(\omega )\exp\left(-{\hbar \omega \over k_BT}\right)
\ \ {\rm (Equilibrium)}.
\end{equation}
The notion of having a noise temperature 
\begin{math} T_n(\omega ) \end{math}
which depends on frequency follows from the definition 
\begin{equation}
S_-(\omega )=S_+(\omega )
\exp\left(-{\hbar \omega \over k_BT_n(\omega )}\right)
\ \ {\rm (General)}.
\end{equation}

The Kubo formula for the electric polarizability of the target, 
\begin{equation}
\alpha (\zeta )=\left({i\over 3\hbar }\right)\int_0^\infty e^{i\zeta t} 
\left<{\bf p}(t)\cdot{\bf p}(0)-{\bf p}(0)\cdot{\bf p}(t)\right>dt, 
\end{equation}
may be written in terms of the spectral functions in Eq.(23) 
employing 
\begin{eqnarray}
&&\left<{\bf p}(t)\cdot{\bf p}(0)-
{\bf p}(0)\cdot{\bf p}(t)\right> \nonumber \\
&= &3\int_{-\infty}^\infty e^{-i\omega t}
\left(S_+(\omega )-S_-(\omega )\right)d\omega .
\end{eqnarray}
Eqs.(27) and (28) imply 
\begin{equation}
\hbar \alpha (\zeta )=\int_{-\infty }^\infty 
\left({S_+(\omega )-S_-(\omega )\over \omega -\zeta }\right) d\omega , 
\end{equation}
from which one may deduce that 
\begin{equation}
{\Im }m\left(\alpha (\omega +i0^+)\right)=
\left({\pi \over \hbar}\right)\left(S_+(\omega )-S_-(\omega ) \right).
\end{equation}
Eqs.(9), (26) and (30) imply  
\begin{equation}
\sigma_{tot}(\omega )=\left({4\pi^2 \omega \over \hbar c}\right)
\left(1-e^{-\hbar \omega /k_BT_n(\omega )}\right)S_+(\omega ).
\end{equation}
One may define the ``symmetrical'' noise spectral function 
according to \begin{math}\bar{S} =(1/2)(S_+ +S_-)\end{math}.  
In terms of  
\begin{equation}
\bar{S}(\omega )={1\over 2}
\left(1+e^{-\hbar \omega / k_BT_n(\omega )}\right)S_+(\omega ),
\end{equation}
which in virtue of Eq.(23) obeys
\begin{equation}
\bar{S}(\omega )\ge 0,
\end{equation}
Eq.(31) reads 
\begin{equation}
\sigma_{tot}(\omega )=\left({8\pi^2 \omega \over \hbar c}\right)
\tanh\left({\hbar \omega \over k_B T_n (\omega )}\right)
\bar{S}(\omega ).
\end{equation}

Eqs.(33) and (34) are the central results of this section. They 
allow Eq.(22) to be written in terms of the noise temperature as 
\begin{equation}
\matrix{
({\rm absorbing})\ T_n(\omega )>0 
\ \Longrightarrow \ (\sigma_{tot}(\omega )>0) \cr 
{\rm and} \cr 
({\rm amplifying})\ T_n(\omega )<0 
\ \Longrightarrow \ (\sigma_{tot}(\omega )<0).  
}
\end{equation}
Thus, the frequency band \begin{math} {\cal B} \end{math} in Eq.(13) 
which is amplifying can be characterized as having a negative noise 
temperature \begin{math} T_n(\omega \in {\cal B})<0 \end{math}. 
As will be discussed in the concluding Sec.V, a variety of 
negative temperature systems\cite{15,16} can be thought to exhibit the 
electromagnetic Klein paradox.

\section{Generalized Optical Theorem}

In this section it is shown why the optical theorem of 
Eq.(8) holds true for an elastic scattering amplitude at 
positive or negative noise temperature. The negative noise 
temperature regime gives rise to a negative total cross section. 

Suppose an electromagnetic wave is incident on a target localized near 
the origin of the spatial coordinate system. The electric field far 
from the target then has the familiar scattering form as 
\begin{math} r\to \infty  \end{math}; i.e.  
\begin{equation}
{\bf E}\to {\Re}e \left\{E_0 e^{-i\omega t}
\left({\bf e}_i e^{i\omega z/c}+
{\bf F\cdot e}_i{e^{i\omega r/c}\over r}\right)
\right\},
\end{equation}
where \begin{math} {\bf F} \end{math} is the dyadic elastic scattering 
amplitude 
\begin{equation}
F_{i\to f}={\bf e}^*_f {\cdot \bf F \cdot}{\bf e}_i .
\end{equation}
The intensity of the incident wave in Eq.(36) is given by 
\begin{equation}
I_0=\left({c|E_0|^2\over 8\pi }\right).
\end{equation}
The intensity of the full wave is given by 
\begin{equation}
I=\left({c\overline{|{\bf E}|^2}\over 4\pi }\right).
\end{equation}

The optical theorem proof envisages detecting radiation 
directly behind the target but far away. We employ polar 
coordinates 
\begin{math} {\bf r}=(x,y,z)=({\bf r}_\perp ,z) \end{math}
and the limits \begin{math} z\to \infty \end{math} so that  
\begin{math} r_\perp << z \end{math}. The total cross section 
is then usually viewed in terms of the ``missing intensity'' 
detected on a screen placed directly behind but far away from 
the target. In mathematical terms, the total cross section may 
be defined as 
\begin{equation}
\sigma_{total}=\lim_{z\to \infty }\int 
\left({I_0-I({\bf r}_\perp,z)\over I_0}\right)d^2{\bf r}_\perp .
\end{equation}

It is crucial to understand the physical meaning of Eq.(40). 
If the intensity behind the screen \begin{math} I  \end{math} 
is less than the incident intensity \begin{math} I_0 \end{math}, 
i.e. if \begin{math} I({\bf r}_\perp,z\to \infty )<I_0 \end{math}, 
then the total cross section is positive 
\begin{math} \sigma_{tot}>0 \end{math} which is most often 
presumed. The intensity behind the target is usually less 
than in the incident radiation because 
(i) the radiation is elastically scattered away at some angle 
with cross section \begin{math} \sigma_{el} \end{math} or 
(ii) the radiation was inelastically absorbed by the target 
with cross section \begin{math} \sigma_{in} \end{math}. 
Altogether,  
\begin{math} \sigma_{tot}=\sigma_{el}+\sigma_{in} \end{math} 
as in Eq.(10). But now consider Eq.(40) if the target is in 
an excited state. The target can be induced to decay by the 
incident radiation making the intensity on the behind the target 
screen bigger than the intensity of the incident radiation; 
i.e. \begin{math} I({\bf r}_\perp,z\to \infty )>I_0 \end{math} 
is a distinct possibility due to the {\em added radiation} 
from the decaying target. The excited target case 
\begin{math} \sigma_{tot}<0 \end{math} is thus quite possible. 

In either of the above cases the optical theorem of Eq.(8) holds 
true as we shall now prove. From Eqs.(36)-(39) it follows that 
as \begin{math} z\to \infty \end{math} and as 
\begin{equation} 
r=\sqrt{z^2+r_\perp ^2}\to z+(r_\perp ^2/2z)+ ...\ , 
\end{equation} 
the intensity on the screen behind the target obeys 
\begin{equation}
\left({I_0-I({\bf r}_\perp,z)\over I_0}\right)\to 
-{\Re }e\left({2F_{i\to i}e^{i(\omega r_\perp ^2/2cz)}\over z}\right),
\end{equation}
which comes from the interference term when one takes the 
absolute square of the amplitude in Eq.(36). Now, 
Eqs.(40) and (42) imply the total cross section 
\begin{equation}
\sigma_{tot}=-\lim_{z\to \infty}{\Re }e\left\{\int 
\left({2F_{i\to i}e^{i(\omega r_\perp ^2/2cz)}\over z}\right) 
d^2{\bf r}_\perp \right\}.
\end{equation}
Using 
\begin{math} 
\int (...) d^2{\bf r}_\perp \to 
2\pi \int_0^\infty (...)r_\perp dr_\perp 
\end{math} 
allows for a simple evaluation of the integral in Eq.(43) leading 
to the optical theorem 
\begin{equation}
\sigma_{tot}=
\left({4\pi c\over \omega }\right){\Im}m \big(F_{i\to i}\big). 
\end{equation}

It is important to note that the proof of the optical theorem 
given above in no manner invokes the sign of the total cross section. 
If the net intensity on the detectors directly behind the target 
is less than the incident intensity, then 
\begin{math} \sigma_{tot}>0 \end{math}. 
If the net intensity on the detectors directly behind the target 
is more than  the incident intensity, then 
\begin{math} \sigma_{tot}<0 \end{math}. In either case Eq.(44) holds 
true.

\section{Conclusions}

We have shown that the electromagnetic Klein paradox arises 
in the form of a possible negative total cross section 
\begin{math} \sigma_{tot} \end{math} for radiation to scatter 
off a target in an excited state. An excited state target can 
be viewed as having a negative noise temperature. Negative noise 
temperature targets are a reality\cite{17,18,19} for LASER or 
MASER pumped materials. 

A negative electromagnetic cross section simply means there is 
more radiation energy behind the target in the outgoing state 
than there would be if the target were {\em not} present. 
The target then represents an electromagnetic amplifier. 
The amplifier must be pumped into an excited state by an external 
energy source before the next incident wave is sent to scatter off 
the target. Such amplifiers have been of considerable interest in 
astrophysics\cite{20,21,22,23,24} where they occur naturally 
in interstellar media. 

The relativistic Klein paradox does not in reality imply any 
sort of negative probability. In the Dirac theory, one may 
create (from pair production) more electrons in the final state 
than were present in the initial state. But in probability terms 
{\em total charge} would remain conserved\cite{25,26,27,28,29,30}. 
Similarly, in the Maxwell theory, one may create (from atomic 
quantum state decays) more photons in the final state than were 
present in the initial state. But in probability terms the 
{\em total energy} would remain conserved. From the above optical 
theorem proof, the total cross section for the scattering of 
electromagnetic radiation may obey 
\begin{math} \sigma_{tot}(\omega \in {\cal B})<0  \end{math} 
because the radiation energy in the final state may exceed (by 
far) the radiation in the initial state, but the probability for 
the scattering event is nevertheless positive. In this manner, 
the electromagnetic Klein paradox may be resolved.

\end{document}